\documentclass[reprint,aip,apl,superscriptaddress]{revtex4-2}

\usepackage{graphicx}
\usepackage{dcolumn}
\usepackage{bm}
\usepackage{physics}
\usepackage{mhchem}
\usepackage{CJKutf8}
\usepackage{empheq}
\usepackage{siunitx}
\usepackage{natbib}
\usepackage{tcolorbox}
\newtcolorbox{mybox}[1]{fonttitle=\bfseries,title=#1}
\newcommand{\ii}{{\rm i}}

\begin{document}

\title{Quantum Correlation and Synchronisation-Enhanced Energy Transfer in Driven-Dissipative Light-Harvesting Dimers}

\author{Wenhao Xu \begin{CJK*}{UTF8}{gbsn}(徐\ 文昊)
\end{CJK*}}
\email{wx266@cam.ac.uk}
\affiliation{Department of Physics and Astronomy, University of Exeter, Exeter EX4 4QL, United Kingdom}
\affiliation{Living Systems Institute, University of Exeter, Stocker Road, Exeter EX4 4QD, United Kingdom}
\affiliation{Brain Physics Laboratory, Department of Clinical Neurosciences, University of Cambridge, Cambridge CB2 0QQ, United Kingdom}

\begin{abstract}
    Quantum synchronisation has recently been proposed as a mechanism for electronic excitation energy transfer in light-harvesting complexes, yet its robustness in driven-dissipative settings remains under active investigation. Here, we revisit this mechanism in cryptophyte photosynthetic antennae using an exciton--vibrational dimer model. By comparing the full open quantum dynamics with semi-classical rate equations for electronic density-matrix elements and vibrational observables, we demonstrate that quantum correlations between electronic and vibrational degrees of freedom, beyond the semi-classical factorised limit, underpin the emergence of quantum synchronisation. Furthermore, we introduce an environment-assisted transfer mechanism arising as a nonlinear, non-Condon correction to the dipole--dipole interaction. This mechanism enables long-lived quantum coherence and continuous, synchronisation-enhanced energy transfer in a driven-dissipative regime, thereby suggesting new avenues for investigating photosynthetic energy-transfer dynamics.
\end{abstract}
\maketitle

\section{Introduction}
Understanding long-lived quantum coherence in photosynthetic energy-transfer dynamics remains a central problem in complex, dissipative molecular environments \cite{cao2020quantum,cheng2009dynamics,christensson2012origin}. In natural photosynthetic systems, incident light is absorbed by chromophores in peripheral antenna complexes, generating electronic excitations that must be efficiently transferred through the complex to a reaction centre to enable the conversion of absorbed solar energy into usable chemical potential \cite{blankenship2021molecular,scholes2011lessons}. Early theoretical descriptions relied on F\"orster resonance energy transfer (FRET), which treats pairwise donor--acceptor coupling perturbatively and assumes rapid environmental decoherence, yielding efficient transfer between chromophores only under near-resonant conditions \cite{tokmakoff2014,fassioli2014photosynthetic}. Subsequent experimental observations of coherent oscillatory signals in photosynthetic antennas, including but not limited to pigment--protein complexes, revealed that electronic couplings can be sufficiently strong to generate delocalised excitons over multiple chromophores \cite{engel2007evidence,zhu2024quantum,hildner2013quantum,panitchayangkoon2010long,brixner2005two,schlau2012elucidation,thyrhaug2018identification,harel2012quantum,panitchayangkoon2011direct}. However, excitonic delocalisation alone cannot account for the persistence of quantum coherence on energy-transfer timescales, prompting growing attention to coupling between electronic excitations and their reaction coordinates, namely intramolecular vibrational modes \cite{chin2013role, anto2021capturing}. Within this context, the exciton--vibrational dimer model is therefore introduced not only as a system-specific approximation, but also as a minimal, homologous two-level system model that isolates the essential dynamical ingredients common to a broad class of light-harvesting and energy-transfer mechanisms \cite{polyutov2012exciton,meskers2023exciton,mukamel2004many,dimitriev2022dynamics}. These developments exposed a fundamental limitation of closed or purely perturbative approaches, thereby motivating the systematic adoption of an open quantum system (OQS) framework, originally developed to describe dissipative system--environment interactions as part of exciton--vibrational intrinsic dynamics \cite{breuer2002theory}. Nevertheless, electronic coherence remains vulnerable to aggressive background decoherence potentially arising from electronic dephasing, intra-molecular vibrations, and solvent phonon coupling in an ambient molecular environment, which could dissipate donor excitation before transfer to the acceptor is completed \cite{pachon2011physical,kreisbeck2012long,nelson2018role,scheidegger2025can,chen2016phonon}. A long-lived coherence capable of sustaining non-trivial energy transfer is therefore empirically suggested, yet its underlying dynamical mechanism remains elusive \cite{reitz2018energy,zhang2016origin,lim2015vibronic}.
\par 
Following this line of inquiry, we re-examine the exciton--vibrational dimer model \cite{schroter2015exciton,womick2011vibronic,womick2009exciton} under both electronic dephasing and vibrational relaxation in the weak-coupling limit. Our analysis focuses on the rate equations for the electronic density-matrix elements and vibrational-mode observables. To ensure physical relevance, we assess both the fully quantum rate-equation formulation and its factorised semi-classical counterpart, parametrised using experimental data from cryptophyte photosynthetic antennae \ce{{PE545}{50/61C}} and \ce{{PE545}{50/61D}} \cite{kolli2012fundamental,doust2004developing,novoderezhkin2010excitation,spear1989unique}. In the absence of electronic dephasing and when only vibrational relaxation is present, we reconfirm in Sec. \ref{sec:results} that quantum synchronisation can facilitate electronic excitation energy transfer, which has recently attracted significant attention \cite{siwiak2019transient,siwiak2020synchronization,zhu2024quantum} in the context of vibronic coupling between donor and acceptor electronic states. More generally, quantum synchronisation is known to emerge in non-equilibrium steady states characterised by limit cycles sustained through continuous system--environment interactions \cite{tao2025noise,cont25,buvca2022algebraic,schmolke2022noise,Dodin_2021}. In photosynthetic complexes, previous studies have theoretically demonstrated phase synchronisation between vibrational displacements at donor and acceptor sites, leading to sustained Pearson correlations \cite{siwiak2019transient,siwiak2020synchronization}. Consistent with this picture, we further show in Sec. \ref{sec:corr} that the quantum synchronisation, quantified using the phase-locking value between vibrational displacement observables, is induced by non-negligible quantum correlations between electronic and vibrational degrees of freedom. These correlations are overlooked under the semi-classical factorised limit, highlighting the intrinsic quantum nature of the synchronisation during the energy-transfer dynamics. In the presence of electronic dephasing, nonetheless, quantum synchronisation does not develop on timescales long enough for energy transfer, but instead emerges only after the electronic excitation has already dissipated at the donor. To render non-trivial energy transfer and sufficiently sustained coherent dynamics under ambient decoherence acting on both electronic and vibrational degrees of freedom, we introduce in Sec. \ref{sec:EAT} an environment-assisted transfer mechanism based specifically on a nonlinear, non-Condon modulation of the static dipole--dipole interaction arising from nuclear normal modes in the residual bath that modulate the donor--acceptor separation \cite{arsenault2021vibronic,anto2021capturing,chin2013role}. When the donor chromophore is continuously driven by coherent light, this driven-dissipative, environment-assisted transfer regime supports sustained energy transfer to the acceptor together with long-lived quantum coherence and synchronisation, thereby offering an alternative and complementary perspective on energy transfer in strongly dissipative, ambient molecular environments.

\section{Theoretical Background\label{sec:theory}}
The historical development of FRET theory reflects both quantum and classical aspects \cite{nelson2018role}. It originates from an electronic two-level system with vibronic degrees of freedom, in which quantised nuclear motions are coupled to the electronic states through a displaced harmonic oscillator model. Within the Born-Oppenheimer approximation, the electronic energies depend only parametrically on vibrational excitation, represented by displacements of the nuclear coordinate $X$ on the potential energy surface \cite{tokmakoff2014,christensson2012origin}. In its final formulation, however, FRET yields an effective classical chemical reaction rate derived using the Fermi golden rule, which implicitly incorporates rapid quantum decoherence. 
\par 
The Hamiltonian of a chromophore can be expressed in terms of product states of electronic and nuclear configurations \cite{knoester2003frenkel,schroter2015exciton}:
\begin{align}
    \begin{split}
        \ket{E}\equiv&\ket{e,n_e}=\ket{e}\otimes\ket{n_e}\\
    \ket{G}\equiv&\ket{g,n_g}=\ket{g}\otimes\ket{n_g}.
    \end{split}
\end{align}
Here, the electronic transition between bound potential energy surfaces can be modelled within a two-level system of $\{\ket{e},\ket{g}\}$, in which an electron is excited to the lowest unoccupied molecular orbital, leaving a hole in the highest occupied molecular orbital. The nuclear configuration is represented by the energy eigenstates $\{\ket{n_e},\ket{n_g}\}$ of harmonic oscillators, each coupled to its corresponding electronic state. Hence, the total system of two chromophores has four basis states: 
\begin{align}
    \{\ket{E_1;E_2}, \ket{E_1;G_2}, \ket{G_1;E_2}, \ket{G_1;G_2}\}.\label{eq:basis}
\end{align}
Each chromophore has an electronic excitation of energy $\epsilon_m$ from a common ground state energy and is strongly coupled to a vibrational mode of frequency $\omega_m$. The free Hamiltonian for the chromophore dimer then reads \cite{zhu2024quantum}
\begin{align}
\begin{split}
    &\hat{H}=\hat{T}+[\frac{1}{2}\omega_1^2\hat{X}_1^2+\frac{1}{2}\omega_2^2\hat{X}_2^2]\op{G_1;G_2}\\
    &+[\epsilon_1+\frac{1}{2}\omega_1^2(\hat{X}_1-d_1)^2+\frac{1}{2}\omega_2^2\hat{X}_2^2]\op{E_1;G_2}\\
    &+[\epsilon_2+\frac{1}{2}\omega_1^2\hat{X}_1^2+\frac{1}{2}\omega_2^2(\hat{X}_2-d_2)^2]\op{G_1;E_2}\\
    &+[\epsilon'+\frac{1}{2}\omega_1^2(\hat{X}_1-d)^2+\frac{1}{2}\omega_2^2(\hat{X}_2-d_2)^2]\op{E_1;E_2}
\end{split} \label{eq:dimer}
\end{align}
where $\hat{T}$ is the kinetic energy operator of the vibrational mode, $\frac{1}{2}\omega_m^2\hat{X}^2$ is the potential energy surface of the $m$th chromophore, and $\hat{X}_m-d_m$ represents the $m$th single excited state displaced by $d_m$ relative to the ground state along the coordinate $X_m$ \cite{tokmakoff2014}. It should be noted that the momentum operator here is essentially $\hat{T}\otimes\hat{1}_{\rm el}$, where $\hat{1}_{\rm el}$ is the identity operator that traces all electronic states. This is because $\hat{T}$ does not directly contribute to the electronic--vibrational coupling. But we adopt $\hat{T}$, as in Ref. \cite{zhu2024quantum}, for notational simplicity. The single electronic excitation is denoted by $\epsilon_m$, which is the shift from the ground state energy $\mel*{G_1;G_2}{\hat{H}}{G_1;G_2}$, and $\epsilon'$ represents the double excitation. 
\par
We model the vibronic Hamiltonian to be harmonic in the both ground and excited states ($\hbar=1$):
\begin{align}
    \hat{H}_{\rm vib}=\omega_1 \hat{b}_1^{\dagger}\hat{b}_1 + \omega_2 \hat{b}_2^{\dagger}\hat{b}_2,
\end{align}
where $\{\hat{b}_{m}^{\dagger},\hat{b}_{m}\}$ are creation and annihilation operators of nuclear vibration. Thus, we can replace the displacement and momentum operators in Eq. \eqref{eq:dimer} by ($m=\{1,2\}$):
\begin{align}
    \hat{X}_m\equiv&\sqrt{\frac{1}{2\omega_m}}(\hat{b}^{\dagger}_m+\hat{b}^{\dagger}_m)&
    \hat{P}_m\equiv&\ii\sqrt{\frac{\omega_m}{2}}(\hat{b}^{\dagger}_m-\hat{b}^{\dagger}_m).
\end{align}
\par 
If we apply the Heitler--London approximation that ignores double excitations $\op{E_1;E_2}$, de-excitations $\op{G_1;G_2}$, and the transitions between them, i.e. $\mel*{E_1;E_2}{\hat{H}}{G_1;G_2}$ \cite{knoester2003frenkel}, then Eq. \eqref{eq:dimer} becomes
\begin{align}
    \begin{split}
    \hat{H}=&\omega_1\hat{b}^{\dagger}_1\hat{b}_1+\omega_2\hat{b}^{\dagger}_2\hat{b}_2\\
    &+ \qty(\epsilon_1+\frac{1}{2}\omega_1^2d_1^2-\omega_1^2\hat{X}_1d_1)\op{E_1;G_2}\\
    &+ \qty(\epsilon_2+\frac{1}{2}\omega_2^2d_2^2-\omega_2^2\hat{X}_2d_2)\op{G_1;E_2}\label{eq:HL}
\end{split}
\end{align}
We hereby introduce the reorganisation energy $\frac{1}{2}\omega_m^2d_m^2=\omega_mS_m$, where $S_m$ denotes the Huang-Rhys factor from the measurements of the Stokes shift. This factor acted on the populations of single excited states can be merged into electronic energies $\varepsilon_m \equiv \epsilon_m+\omega_mS_m$. The electronic--vibrational interactions stem from the second and third lines of Eq. \eqref{eq:HL}.
\begin{figure}[t]
    \centering
    \includegraphics[width=0.8\linewidth]{img/Schematics.pdf}
    \caption{Schematic illustration of an exciton–vibrational dimer weakly coupled to its environment, including vibronic relaxation and electronic dephasing channels with rates $\Gamma_{\rm th}$ and $\Gamma_{\rm deph}$, receptively. The additional rate $\Gamma_{\rm dip}$ denotes the introduced environment-assisted contribution to the dipole--dipole interaction.}
    \label{fig:scheme}
\end{figure}
\par 
In a molecular dimer, dipole--dipole coupling is considered to exchange electronic excitations within the point-dipole approximation, where permanent dipole moments of the molecular ground and excited states are neglected. Within FRET, the Condon approximation is typically employed, assuming that electronic transitions occur on timescales much shorter than nuclear motion, such that the nuclear dependence of the electronic transition dipole moment can be neglected \cite{tokmakoff2014}:
\begin{align}
    \hat{V}_{\rm dip}=J(\op{2}{1}+\op{1}{2}),
\end{align}
where $\ket{1}\equiv\ket{E_1;G_2}$ and $\ket{2}\equiv\ket{G_1;E_2}$ are denoted, and the coupling strength is denoted by $J$. In Sec. \ref{sec:EAT}, we go beyond the Condon approximation.
\par 
The aforementioned delocalised electronic states occur when the electronic coupling is non-negligible compared to the energy splitting between the states ($\epsilon_1-\epsilon_2$). However, this coupling strength is still much smaller than the
electronic excitations $(J\ll\epsilon_2<\epsilon_1\ll\epsilon')$, allowing us to focus on the single-excited states while assuming trivial fillings in the doubly (de)excited states. The effective system Hamiltonian on the two-chromophore basis therefore consists of single electronic excitations, dipole--dipole couplings between the single excited states, vibronic oscillators, and electronic couplings to vibrational modes:
\begin{align}
\begin{split}
    \hat{H}_{\rm S}=&\frac{\Delta \varepsilon}{2}\qty(\hat{\sigma}_1^+\hat{\sigma}_1^--\hat{\sigma}_2^+\hat{\sigma}_2^-) + J\qty(\hat{\sigma}_1^+\hat{\sigma}_2^-+\hat{\sigma}_2^+\hat{\sigma}_1^-)\\
    &+\omega_{1}\hat{b}_{1}^{\dagger}\hat{b}_{1} + \omega_{2}\hat{b}_{2}^{\dagger}\hat{b}_{2}\\
    &+g_1\op{1}(\hat{b}_{1}^{\dagger}+\hat{b}_{1})+g_2\op{2}(\hat{b}_{2}^{\dagger}+\hat{b}_{2}).
\end{split}\label{eq:model}
\end{align}
where the electronic coupling strength to its vibrational mode is denoted by $g_m\equiv-\omega_m\sqrt{S_m}$. Choosing an appropriate gauge between the electronic energies, we can concentrate on the energy difference $\Delta\varepsilon\equiv\varepsilon_1-\varepsilon_2$. We can further convert $\op{m}$ to spin operators: $\op{1}=\hat{\sigma}_1^+\hat{\sigma}_1^-(\hat{1}-\hat{\sigma}_2^+\hat{\sigma}_2^-)$ and $\op{2}=(\hat{1}-\hat{\sigma}_1^+\hat{\sigma}_1^-)\hat{\sigma}_2^+\hat{\sigma}_2^-$. But for notation simplicity, we keep $\op{m}$.
\par 
The electronic Hamiltonian \eqref{eq:model} can be further reduced under the basis $\{\ket{1},\ket{2}\}$ for single excited states only (like in Ref. \cite{zhu2024quantum,o2014non}). The electronic Hamiltonian is of the standard Jaynes-Cummings form:
\begin{align}
    \hat{H}_{\rm el}=&\frac{\Delta \varepsilon}{2}\hat{\sigma}_z + J\hat{\sigma}_x.\label{eq:9}
\end{align}
Its eigenstates are the so-called one-exciton states, which are symmetric and antisymmetric combination of donor's and acceptor's single excited states: $\ket{\pm} = (\ket{1}\pm\ket{2})/\sqrt{2}$ with eigenvalues $E_{\pm} = \pm\sqrt{(\Delta\varepsilon/2)^2+J^2}$ \cite{knoester2003frenkel,schroter2015exciton}.
Clearly, $\ket{\pm}$ is delocalised across two chromophores. One might notice that in Eq. \eqref{eq:9}, the correspondences $\hat{\sigma}_z\Leftrightarrow\hat{\sigma}_1^+\hat{\sigma}_1^--\hat{\sigma}_2^+\hat{\sigma}_2^-$ and $\hat{\sigma}_x\Leftrightarrow\hat{\sigma}_1^+\hat{\sigma}_2^-+\hat{\sigma}_2^+\hat{\sigma}_1^-$ are substituted from Eq. \eqref{eq:model}. However, we need to model electronic dephasing of population states $\op{1}$ and $\op{2}$ carried by $\hat{\sigma}_m^-$. This forces us to adhere to the two-chromophore basis \eqref{eq:basis}, since the $\hat{\sigma}_m^-$ dephasing transitions the populations of single excitations from/to the doubly (de)excited states.
\par
Although the couplings between the electronic and vibrational states are considered strong, their couplings to environment can be taken to be weak \cite{siwiak2019transient,siwiak2020synchronization}. In this weak-coupling limit, the system's equation of motion yields the Lindblad master equation with eight possible collapse operators $\hat{C}_{\nu}=\{\hat{\sigma}^{\pm}_m,\hat{b}_m,\hat{b}_m^{\dagger}\}$ with $m=\qty{1,2}$ \cite{breuer2002theory}:
\begin{align}
    \begin{split}
        &\dv{}{t}\hat{\rho}(t)=-\ii [\hat{H}_{\rm S},\hat{\rho}(t)]\\
        +&\sum_{\nu}\Gamma_{\nu}\qty(\hat{C}_{\nu}\hat{\rho}(t)\hat{C}_{\nu}^{\dagger}-\frac{1}{2}\hat{C}_{\nu}^{\dagger}\hat{C}_{\nu}\hat{\rho}(t)-\frac{1}{2}\hat{\rho}(t)\hat{C}_{\nu}^{\dagger}\hat{C}_{\nu})
    \end{split}
    \label{eq:froster_eom}
\end{align}
in an electronic dephasing rate $\Gamma_{\rm deph}$ and a vibronic dissipation rate $\Gamma_{\rm th}$. The weak-coupling limit requires that $\Gamma_{\nu}\ll\Delta\varepsilon/\hbar$ and $\Gamma_{\nu}\ll\omega_m$ hold. 
\par 
Note that the contributions by $\hat{\sigma}^{+}_m$ and $\hat{b}^{\dagger}_m$ are negligible, because they can hardly raise the system state by annihilating an environmental phonon in the rate $\Gamma_{\nu}\bar{n}_{\rm ph}$, where $\bar{n}_{\rm ph}=(e^{\beta \omega_{\nu}}-1)^{-1}$ denotes the mean number of quanta in a thermally occupied mode at inverse temperature $\beta$. When the transition frequency of phonon energy quanta matches with electronic energy gap $\omega_{\nu}=\Delta\varepsilon_m/\hbar$ (or vibrational energy quanta $\omega_{\nu}=\omega_m$), $\bar{n}_{\rm ph}\approx0$. Hence, $\hat{\sigma}^{+}_m$ and $\hat{b}^{\dagger}_m$ will have trivial effects in zero rates and not be considered (For details, please refer to the SM). The whole scheme can be illustrated in Fig. \ref{fig:scheme}
\par 
We attempt to analyse wave-like behaviour predicted in the RET theory via rate equations of population states and corresponding physical quantities. The rate equations of interest can be obtained by taking the expectation value of $\op{m}$, $\hat{\sigma}_1^+\hat{\sigma}_2^-$, $\hat{\sigma}_2^+\hat{\sigma}_1^-$, $\hat{b}_m^{\dagger}\pm\hat{b}_m$ on Eq. \eqref{eq:froster_eom} with four effective collapse operators $\{\hat{\sigma}_m^-,\hat{b}_m\}$:
\begin{widetext}
\begin{subequations}
\begin{eqnarray}
    \dv{}{t}\rho_{11}&=&-\ii J(\expval*{\hat{\sigma}_1^+\hat{\sigma}_2^-}-\expval*{\hat{\sigma}_2^+\hat{\sigma}_1^-})-\Gamma_{\rm deph}(\rho_{11}-\rho_{00})\\
    \dv{}{t}\rho_{22}&=&-\ii J(\expval*{\hat{\sigma}_2^+\hat{\sigma}_1^-}-\expval*{\hat{\sigma}_1^+\hat{\sigma}_2^-})-\Gamma_{\rm deph}(\rho_{22}-\rho_{00})\\
    \dv{}{t}\expval*{\hat{\sigma}_1^+\hat{\sigma}_2^-}&=&\ii\Delta\varepsilon\expval*{\hat{\sigma}_1^+\hat{\sigma}_2^-}+\ii J(\rho_{11}-\rho_{22})+\ii g_1\expval*{\hat{\sigma}_1^+\hat{\sigma}_2^-(\hat{b}_1^{\dagger}+\hat{b}_1)}-\ii g_2\expval*{\hat{\sigma}_1^+\hat{\sigma}_2^-(\hat{b}_2^{\dagger}+\hat{b}_2)}-\Gamma_{\rm deph}\expval*{\hat{\sigma}_1^+\hat{\sigma}_2^-}\label{eq:corr1}\\
    \dv{}{t}\expval*{\hat{\sigma}_2^+\hat{\sigma}_1^-}&=&-\ii\Delta\varepsilon\expval*{\hat{\sigma}_2^+\hat{\sigma}_1^-}-\ii J(\rho_{11}-\rho_{22})-\ii g_1\expval*{\hat{\sigma}_2^+\hat{\sigma}_1^-(\hat{b}_1^{\dagger}+\hat{b}_1)}+\ii g_2\expval*{\hat{\sigma}_2^+\hat{\sigma}_1^-(\hat{b}_2^{\dagger}+\hat{b}_2)}-\Gamma_{\rm deph}\expval*{\hat{\sigma}_2^+\hat{\sigma}_1^-}\label{eq:corr2}\\
    \dv{}{t}\expval*{\hat{b}_1^{\dagger}+\hat{b}_1}&=&\ii \expval*{\hat{b}_1^{\dagger}-\hat{b}_1}-\frac{\Gamma_{\rm th}}{2}\expval*{\hat{b}_1^{\dagger}+\hat{b}_1}\\
    \dv{}{t}\expval*{\hat{b}_2^{\dagger}+\hat{b}_2}&=&\ii \expval*{\hat{b}_2^{\dagger}-\hat{b}_2}-\frac{\Gamma_{\rm th}}{2}\expval*{\hat{b}_2^{\dagger}+\hat{b}_2}\\
    \dv{}{t}\ii\expval*{\hat{b}_1^{\dagger}-\hat{b}_1}&=& -\expval*{\hat{b}_1^{\dagger}+\hat{b}_1}-2g_1\rho_{11}-\frac{\Gamma_{\rm th}}{2}\ii\expval*{\hat{b}_1^{\dagger}-\hat{b}_1}\\
    \dv{}{t}\ii\expval*{\hat{b}_2^{\dagger}-\hat{b}_2}&=& -\expval*{\hat{b}_2^{\dagger}+\hat{b}_2}-2 g_2\rho_{22}-\frac{\Gamma_{\rm th}}{2}\ii\expval*{\hat{b}_2^{\dagger}-\hat{b}_2},
    \end{eqnarray}\label{eq:system}
\end{subequations}
where $\rho_{00}\equiv\expval{\op{E_1;E_2}}$, $\rho_{11}\equiv\expval{\op{1}}$ and $\rho_{22}\equiv\expval{\op{2}}$ are defined (For derivation details, please see SM). 
\end{widetext}
Recall that the Heitler--London approximation allows us to ignore $\rho_{00}\equiv\expval*{\op{E_1;E_2}}=0$, which is shown to vanish in Fig. (\ref{fig:1}) in the absence of external driving. Eq. \eqref{eq:system} cannot be exactly solved, as the electronic and vibrational states are correlated in the expectation in Eq. \eqref{eq:corr1} and \eqref{eq:corr2}. Thus, we have to analyse the rate equations \eqref{eq:system} through numerical means.
\begin{table}[hbtp]
\caption{\label{tab:1} Parameters used for numerical simulations in this work. All unlabelled units are in spectroscopic wave numbers (cm$^{-1}$). Data is adopted from Ref. \cite{kolli2012fundamental,doust2004developing,novoderezhkin2010excitation}. Simulations of the rate equations \eqref{eq:system} use \texttt{QuTiP} package in \texttt{Python}.}
\begin{ruledtabular}
\begin{tabular}{ccccccccc}
$\Delta\varepsilon$ & $J$ & $k_B T$ & $\omega_1$& $\omega_2$ & $S_1$ & $S_2$ & $\Gamma_{\rm th}$ & $\Gamma_{\rm deph}$\\
\hline
1042 & 92 & 207.1 & 1450 & 1111 & 0.1013 & 0.0578 & 1 THz & 10 THz\\
\end{tabular}
\end{ruledtabular}
\end{table}

\section{Quantum Synchronisation and Energy Transfer\label{sec:results}}
\begin{figure}[t]
    \centering
    \includegraphics[width=\linewidth]{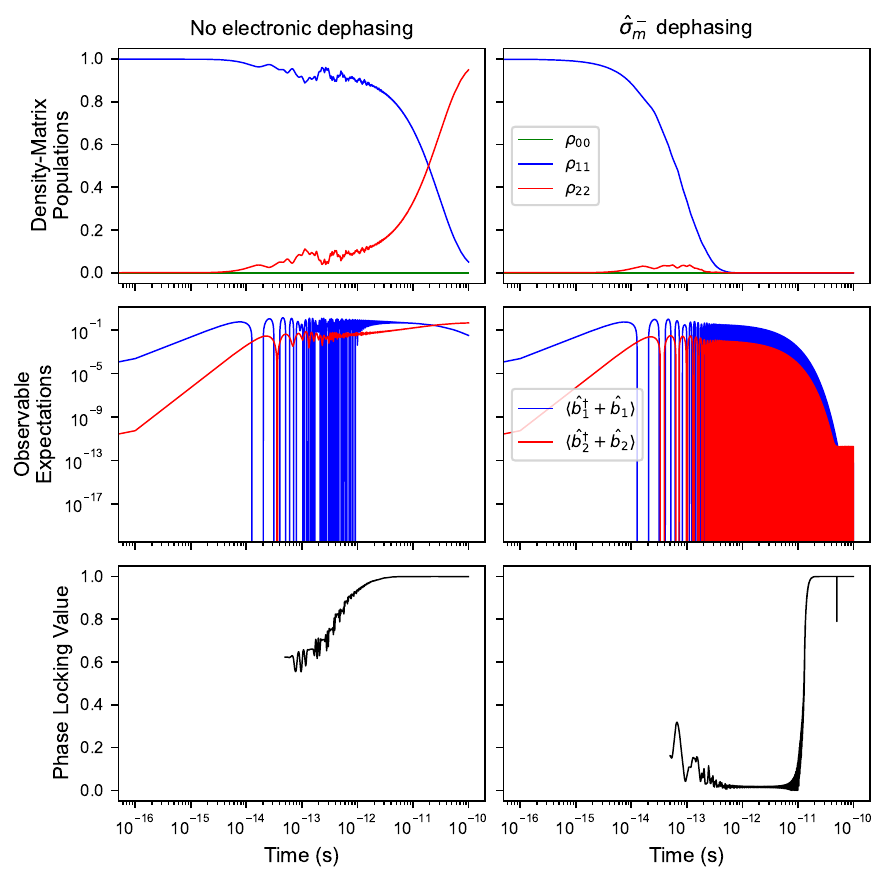}
    \caption{Numerical simulation of Eq. \eqref{eq:froster_eom} under various conditions. The left column of subplots considers the case with vibronic dissipation only, while the right one further includes $\hat{\sigma}_m$ electronic dephasing. The first row depicts the density matrix populations of electronic states. The second row plots $\expval*{\hat{b}_m^{\dagger}+\hat{b}_m}$, and the last row shows the moving PLV of a time window 0.1 ps over the 100 ps simulation. The negative values of $\expval*{\hat{b}_m^{\dagger}+\hat{b}_m}$ have been discarded for logarithmic scales on the vertical axes. Parameters used in the simulation can be found in Table. (\ref{tab:1}).}
    \label{fig:1}
\end{figure}
\par 
Quantum synchronisation in the photosynthesis literature refers to phase alignment of oscillations of vibronic displacement expectations, $\expval*{\hat{b}_1^{\dagger}+\hat{b}_1}$ and $\expval*{\hat{b}_2^{\dagger}+\hat{b}_2}$. Here, we focus on these dimensionless and un-normalised observables for computational simplicity. Previous studies have employed the Pearson correlation and quantum mutual information to characterise the onset of synchronisation \cite{siwiak2020synchronization,siwiak2019transient}. In this work, we instead use the phase-locking value (PLV), which provides a simple, direct measure for real-valued signals,
\begin{align} 
    {\rm PLV}=\abs{\frac{1}{N}\sum_t^Ne^{\ii \Delta\phi(t)}},
\end{align}
where $\Delta\phi(t) = \phi_1(t)-\phi_2(t)$ denotes the instantaneous phase difference between the two signals, and $N$ is the number of data points in the time series. The phases $\phi_m (t)$ are obtained using the Hilbert transform \cite{aydore2013note}. The PLV is a non-negative quantity ranging from 0 to 1, with PLV$=0$ indicating no synchronisation and PLV$=1$ corresponding to perfect phase synchronisation. To track synchronisation dynamically, we apply a moving average with a 0.1 ps window over a 100 ps simulation. All simulations use realistic parameters from cryptophyte photosynthetic antennae PE545$_{\text{50/61C}}$ and PE545$_{\text{50/61D}}$, listed in Table (\ref{tab:1}) \cite{kolli2012fundamental, doust2004developing, novoderezhkin2010excitation, spear1989unique}.
\par 
In the absence of electronic dephasing via $\hat{\sigma}_m^{-}$, but including vibrational dissipation through $\hat{b}_m$, the left column of Fig. (\ref{fig:1}) shows that energy transfer coincides with the emergence of quantum synchronisation. The PLV approaches unity at approximately 4 ps, indicating synchronisation between $\expval*{\hat{b}_1^{\dagger}+\hat{b}_1}$ and $\expval*{\hat{b}_2^{\dagger}+\hat{b}_2}$. Although the acceptor population $\rho_{22}$ begins  to increase slightly around this time, the dominant energy transfer from molecule 1 to molecule 2 occurs subsequently, concomitant with sustained synchronised dynamics. This behaviour reflects population conservation via vibronic oscillations, where excitation is completely transferred from $\rho_{11}$ to $\rho_{22}$ in the absence of electronic dephasing. 
\par 
When both electronic dephasing via $\hat{\sigma}_m^-$ and vibrational dissipation are present, the right column of Fig. (\ref{fig:1}) shows a different behaviour. In this case, $\rho_{22}$ fails to reach a significant value before dissipation, indicating the suppression of effective energy transfer. Even though the PLV eventually reaches unity at later time, this synchronisation arises from the trivial phase locking between $\expval*{\hat{b}_m^{\dagger}+\hat{b}_m}$ in the steady state after the electronic excitation has already been fully dephased.
\par 
These results extend and contextualises the observations reported in Ref. \cite{siwiak2020synchronization}. There, the Pearson correlation between vibrational displacements approaches unity, indicating near-perfect synchronisation after approximately 8 ps when $\omega_1/\omega_2=1.005$, with the onset of quantum synchronisation shifting to later times as the frequency mismatch increases. In our simulations, the vibrational expectation values $\expval*{\hat{b}_1^{\dagger}+\hat{b}_1}$ and $\expval*{\hat{b}_2^{\dagger}+\hat{b}_2}$ similarly exhibit synchronised dynamics at later times for larger frequency detuning, for example around 15 ps when $\omega_1/\omega_2\approx1.3$, in quantitative agreement with Ref. \cite{siwiak2020synchronization}. However, once electronic dephasing is included, the acceptor population $\rho_{22}$ has already dissipated before synchronisation develops. This implies that quantum synchronisation remains a robust dynamical enhancement to the photosynthetic energy transfer, but that ambient dissipation can obscure its contribution by delaying its onset beyond the timescale of electronic population dynamics, thereby motivating the introduction of additional possible dynamical mechanisms that can sustain coherence and synchronisation over energy-transfer timescales.

\section{Non-negligible Quantum Correlations\label{sec:corr}}
\begin{figure}[t]
    \centering
    \includegraphics[width=\linewidth]{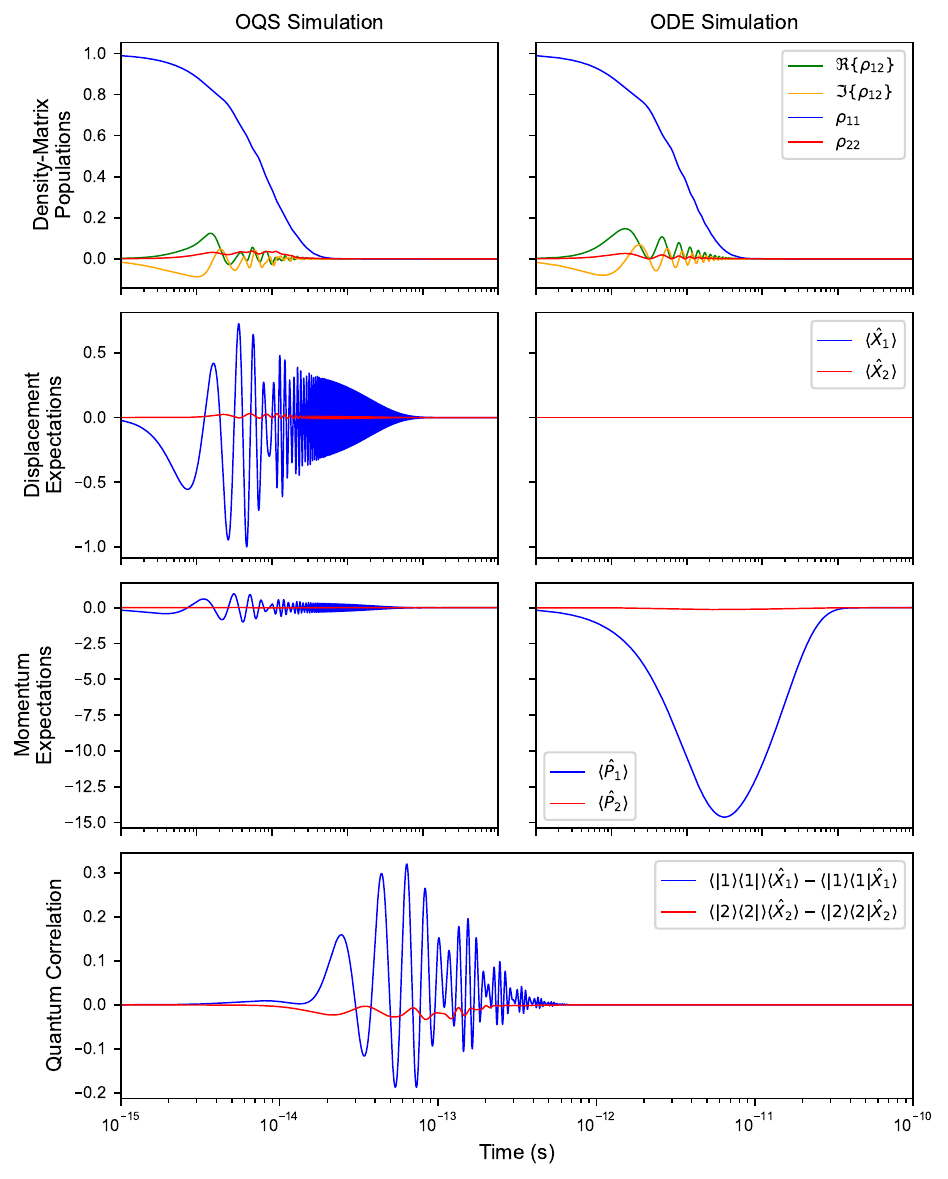}
    \caption{Numerical simulation of the quantum dynamics (\ref{eq:system}) with $\hat{\sigma}_m$ electronic dephasing compared to classical ODE simulation. The first row shows electronic density matrix elements. The second and third rows depict $\expval*{\hat{b}_m^{\dagger}+\hat{b}_m}$ and $\ii\expval*{\hat{b}_m^{\dagger}-\hat{b}_m}$, respectively. The last row plots the quantum correlations $\Delta C_m\equiv \expval*{\op{m}}\expval*{\hat{b}_m^{\dagger}+\hat{b}_m}-\expval*{\op{m}(\hat{b}_m^{\dagger}+\hat{b}_m)}$ due to the factorised-state approximation \eqref{eq:factorized}. Parameters used can be found in Table. (\ref{tab:1}).}
    \label{fig:2}
\end{figure}
We have examined the dissipative influences on the energy-transfer process; however, the quantum nature underlying the synchronising oscillations remains unclear. In this section, we identify non-negligible quantum correlations that mix electronic and vibrational degrees of freedom and mediate the transfer of donor electronic excitation through donor-acceptor vibronic synchronisation down to the acceptor electronic state.
\par 
To elucidate this point by contradiction, we assume that the mixing between electronic and vibrational states is negligible. Under this assumption, the factorised-state approximation,
\begin{align}
\begin{split}
    \expval*{\hat{\sigma}_1^+\hat{\sigma}_2^-(\hat{b}_m^{\dagger}+\hat{b}_m)}&\approx\expval*{\hat{\sigma}_1^+\hat{\sigma}_2^-}\expval*{\hat{b}_m^{\dagger}+\hat{b}_m}\\
    \expval*{\hat{\sigma}_2^+\hat{\sigma}_1^-(\hat{b}_m^{\dagger}+\hat{b}_m)}&\approx\expval*{\hat{\sigma}_2^+\hat{\sigma}_1^-}\expval*{\hat{b}_m^{\dagger}+\hat{b}_m},
\end{split}\label{eq:factorized}
\end{align}
allows the quantum dynamics to be simulated in a semi-classical way using coupled ordinary differential equations (ODEs). In this representation, the system exhibits signatures characteristic of classical, nonlinear stochastic dynamics. Throughout this section, we restrict our analysis to the conventional case including vibronic dissipation and electronic dephasing via $\hat{\sigma}_m^-$ only.
\par 
Fig. (\ref{fig:2}) compares the results obtained from the open quantum system (OQS) and ODE simulations, highlighting the non-classical nature of energy transfer in the dimer model \eqref{eq:model}. The first row of subplots shows close agreement between the two approaches for the populations $\rho_{11}$ and $\rho_{22}$, as well as for the off-diagonal element $\rho_{12}$. In contrast, the second and third rows reveal that, under the factorised-state approximation, the expected vibronic displacements $\expval*{\hat{b}_m^{\dagger}+\hat{b}_m}$ and momenta $\ii\expval*{\hat{b}_m^{\dagger}-\hat{b}_m}$ lack non-trivial oscillatory behaviour. The final row further examines the difference between factorised and unfactorised treatments, showing pronounced deviations for the donor chromophore. These discrepancies indicate the presence of non-negligible quantum correlations between off-diagonal electronic density-matrix elements and vibronic displacements, which lead to appreciable deviations in $\expval*{\hat{b}_m^{\dagger}+\hat{b}_m}$ (and $\ii\expval*{\hat{b}_m^{\dagger}-\hat{b}_m}$) prior to vibronic dissipation.
\par 
In other words, under the factorised-state approximation, vibronic synchronisation is suppressed, with $\rho_{22}$ symmetrically oscillating around zero, such that no net energy transfer can be initiated. When electronic--vibrational correlations are retained, $\rho_{22}$ instead does develop a tiny but finite transient increase. This difference becomes more evident when electronic dephasing via$\hat{\sigma}_m^-$ is switched off, as shown in Fig. (1) of the Supplementary Material. These observations demonstrate that, in the presence of electronic dephasing, non-negligible quantum correlations between electronic and vibrational degrees of freedom play a crucial role in enabling quantum synchronisation and facilitating energy transfer.

\section{Environment Assisted Transfer}\label{sec:EAT}
\begin{figure}[t]
    \centering
    \includegraphics[width=\linewidth]{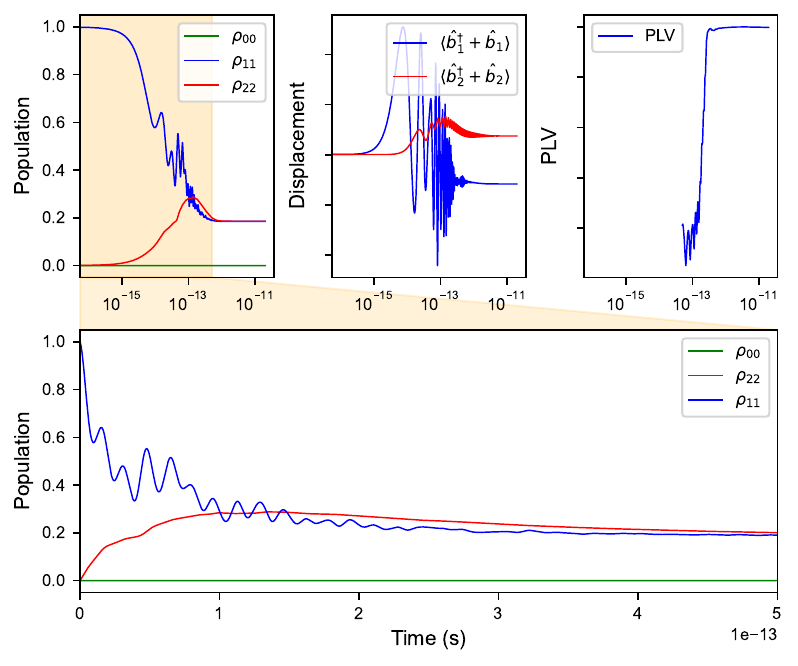}
    \caption{Numerical simulation of the environment-assisted energy transfer process. The first row shows the electronic populations of the single excited states, expected displacements $\expval*{\hat{b}_m^{\dagger}+\hat{b}_m}$, and corresponding PLVs, from left to right, respectively. The second row zooms in on the electronic populations during the first 0.5 ps. Instead of using $\Delta\varepsilon$ in Table. (\ref{tab:1}), we use real site energies, $\varepsilon_1=19574$ and $\varepsilon_2=18532$, assume the double excitation energy to be 60000, which is large enough to keep the Heitler--London approximation valid ($\rho_{00}=0$) and allows us to focus only on the single excited states (all in units of cm$^{-1}$). The Rabi frequency is taken to be $\Omega=200$ THz, the environment-assistant rate is $\Gamma_{\rm dip} = 10$ THz, and the electronic dephasing rate is now $\Gamma_{\rm deph} = 0.1$ THz.}
    \label{fig:3}
\end{figure}
To enhance the efficiency of energy transfer in the presence of both electronic dephasing and vibronic dissipation, we introduce an environment-assisted transfer mechanism in which the electronic coupling is dynamically modulated by environmental degrees of freedom, here taken to be nuclear normal modes. \cite{arsenault2021vibronic,chin2013role,chen2016phonon}. Within this framework, the dipole--dipole interaction no longer act as an independent coherent channel; instead, its effectiveness is also shaped by weak-coupling to environment, which can open additional transfer pathways under driven, dissipative conditions.
\par
Following a non-Condon treatment \cite{arsenault2021vibronic,chin2013role,sun2016non,kundu2022franck}, we expand the dipole--dipole coupling in the normal-mode nuclear coordinates $Q_i$ about the equilibrium configuration,
\begin{align}
    \hat{V}_{\rm dip}(Q) =& \hat{J}_0 + \sum_i\qty(\pdv{\hat{J}}{Q_i})_{Q_i=0}\hat{Q}_i\label{eq:expansion}\\
    \begin{split}
        =&J\qty(\hat{\sigma}_1^+\hat{\sigma}_2^-+\hat{\sigma}_2^+\hat{\sigma}_1^-) \\&+ \sum_i C_i\qty(\hat{\sigma}_1^+\hat{\sigma}_2^-+\hat{\sigma}_2^+\hat{\sigma}_1^-)(\hat{B}_i^\dagger+\hat{B}_i).
    \end{split}
\end{align}
Here, the coordinates $Q_i$ denote nuclear normal modes that modulate the donor-acceptor separation relevant to the electronic coupling, and $\{\hat{B}_i^\dagger, \hat{B}_i\}$ represent the corresponding quantised nuclear creation and annihilation operators operators. Conceptually, this partition of nuclear degrees of freedom is analogous to the reaction-coordinate framework, in which a small number of vibrational modes are retained explicitly in the system Hamiltonian, here referred to as vibrons $\{\hat{b}_m^\dagger, \hat{b}_m\}$ with $m\in\qty{1,2}$, while the remaining nuclear modes are treated as a residual bath \cite{anto2021capturing}. While the equilibrium coupling scales as $\hat{J}_0\sim r^{-3}$, its leading linear correction for modes that modulate the donor-acceptor separation exhibits a leading distance dependence $(\partial\hat{J}/Q_i)_{0}\hat{Q}_i\sim r^{-4}\hat{Q}_i$. 
\par 
Retaining only energy-conserving terms under the rotating-wave approximation, and imposing the resonance condition $\Delta\varepsilon\simeq \Omega_i$ implicit in the Fermi golden rule, the non-Condon interaction Hamiltonian reduces, in the interaction picture, to
\begin{align}
    \delta\hat{J}_I = \sum_i C_i\hat{\sigma}_1^-\hat{\sigma}_2^+\hat{B}_i^{\dagger}e^{-\ii(\varepsilon_1-\varepsilon_2-\omega_i)}+\text{h.c.}.
\end{align}
At ambient temperature, the relevant frequencies and the electronic gap satisfy $\hbar\Omega_i\sim\Delta\varepsilon\gg k_BT$, such that the thermal Bose occupation $\bar{n}_{\rm B}(\Delta\varepsilon)=(e^{\beta\Delta\varepsilon}-1)^{-1}$ is vanishing and $\ll 1$ (here $\bar{n}_{\rm B}\sim 10^{-3}$ for our parameters in Table. \ref{tab:1}). Consequently, non-Condon nuclear relaxation is dominated by emission rather than absorption, allowing the environment-assisted channel to be represented effectively by the dissipative process governed by a single collapse operator $\hat{\sigma}_1^-\hat{\sigma}_2^+$ with factor $(\bar{n}_{\rm B}+1)$. 
\par 
We model the environment-assisted energy transfer using a Lindblad rate $\Gamma_{\rm dip}= 1$ THz $\sim\Gamma_{\rm deph}$, which irreversibly transfers the electronic excitation energy into the nuclear normal modes. This rate corresponds to a moderate modulation of static dipole--dipole coupling, consistent with $\delta J/J_0\sim 3 \delta r/r<1$, thereby ensuring the validity of the expansion \eqref{eq:expansion} while enabling non-trivial population transfer remains efficient on sub-picosecond timescales and preserving electronic coherence in line with experimental observations \cite{richards2012coherent}.
\par 
We next consider a beyond-laboratory, driven-dissipative scenario, illustrated schematically in Fig. \ref{fig:scheme}, in which an outer donor chromophore with higher electronic energy is continuously driven by incident light and transfers excitation energy to inner acceptor chromophores that couple only weakly to the light field. Figure \ref{fig:3} presents the open quantum dynamics in such a configuration, where only chromophore 1 (the donor) is directly driven by coherent light. The light-matter interaction is modelled using the standard resonance-fluorescence Hamiltonian \cite{breuer2002theory},
\begin{align}
    \hat{H}_{\rm d}=\frac{\Omega_{\rm p}}{2}(\hat{\sigma}^+_1e^{-\ii\omega_{\rm p}t}+\hat{\sigma}^-_1e^{\ii\omega_{\rm p}t})
\end{align}
The pumping frequency $\omega_{\rm p}=\omega_1+\epsilon_1/\hbar$ is chosen to resonantly address donor transition, and a large Rabi frequency $\Omega_{\rm p}=200$ THz ensures efficient excitation. In the absence of environment-assisted transfer (cf. the right column of Fig. \ref{fig:1}), the static dipole--dipole coupling $\hat{J}_0$ in PE545 is insufficient to sustain excitation transfer to the acceptor in the presence of $\hat{\sigma}_m^-$ electronic dephasing; as a result, the acceptor population $\rho_{22}$ remains negligible. In contrast, once this $\hat{\sigma}_1^-\hat{\sigma}_2^+$ environment-assisted transfer channel is activated, the upper panel of Fig. \ref{fig:3} shows that $\rho_{22}$ approaches a non-trivial steady-state value. Simultaneously, the PLV rises sharply at the onset of population transfer and remains close to unity in the driven–dissipative steady state. The lower panel of Fig. \ref{fig:3} further reveals pronounced wave-like oscillations in $\rho_{11}$, reminiscent of dynamical features observed experimentally \cite{turner2012quantitative}. Taken together, these observations indicate that quantum synchronisation co-emerges with sustained energy flow in the environment-assisted regime, consistent with long-lived electronic–vibrational coherence and residual quantum correlations induced by nuclear modulation of the electronic coupling.

\section{Conclusion and Outlook}
We have re-examined the exciton--vibrational dimer model in an open quantum formalism within the weak-coupling limit, using rate equations for electronic density-matrix elements together with vibrational displacement observables. Parametrising the model with data motivated by cryptophyte antenna complexes (\ce{{PE545}_{50/61C}} and \ce{{PE545}_{50/61D}}), we compared the fully quantum dynamics with a factorised semi-classical counterpart. This comparison isolates the role of electronic–vibrational state mixing and shows that non-negligible quantum correlations influence vibronic displacement dynamics and induces the synchronisation as well as energy transfer.
\par 
Our results also clarify how synchronisation should be interpreted in the driven-dissipative regime. We quantify synchronisation using the PLV, a standard classical signal-analysis metric applied to the instantaneous phase of the expectation values of vibrational displacement observables. In this setting, PLV provides a simple, straightforward diagnostic of phase alignment of vibronic motion, while remaining agnostic to the microscopic origin of the synchronisation \cite{siwiak2020synchronization,siwiak2019transient,zhu2024quantum}. When the environment-assisted transfer channel and continuous pumping sustain non-trivial energy flow, the PLV rises meaningfully and sharply at the onset of population transfer and remains close to unity in the steady state, concomitant with long-lived coherence and residual electronic--vibrational correlations. These observations support the interpretation that quantum synchronisation plays a central role in energy transfer, with its inherently quantum character manifested in the breakdown of the factorised closure \eqref{eq:factorized}, whereby electronic and vibrational dynamics become mutually conditioned through vibronic state mixing beyond the semi-classical factorised limit, so that neither subsystem can be described as evolving under an independent mean field of the other. At the same time, PLV must be interpreted with care, particularly in the presence of electronic dephasing. When no appreciable excitation can be transferred and the relevant displacement expectations relax into a trivial late-time regime, PLV can still approach unity due to phase estimates becoming effectively locked despite the absence of meaningful synchronised oscillations. We therefore regard PLV as intuitive but only physically significant when assessed together with non-trivial steady-state electronic populations, which distinguishes dynamical synchronisation accompanying transfer from trivial steady-state locking.
\par 
To place these findings in an experimentally accessible context, we emphasise how our model observables correspond to quantities routinely accessed in ultrafast spectroscopy. The electronic populations $\rho_{11}$ and $\rho_{22}$ could be reflected in donor- and acceptor-associated transient-absorption features, since ground-state bleach reports depletion of ground-state population and excited-state absorption requires population in an excited state; time-resolved fluorescence likewise tracks excited-state occupation up to proportionality factors \cite{berera2009ultrafast}. Beyond populations, quantum correlations from electronic--vibrational state mixing can manifest experimentally as oscillatory beat components and cross-peak dynamics in two-dimensional electronic spectroscopy (2DES). In molecular aggregates and light-harvesting complexes, such oscillatory signals in 2DES have been linked to vibrational or vibronic coherence dynamics \cite{milota2013vibronic}, including long-lived beatings arising from exciton--vibrational coupling when vibrational frequencies are resonant with electronic energy gaps \cite{lim2015vibronic}, and characteristic phase behaviour that distinguishes vibronic contributions from purely electronic coherence \cite{perlik2014distinguishing}. Such vibronic beatings, observed in multiple light-harvesting complexes, are widely interpreted as signatures of electronic--vibrational coherence rather than purely electronic excitonic dynamics \cite{engel2007evidence}. The vibrational displacement observables $\expval*{\hat{b}_m^\dagger+\hat{b}_m}$ correspond physically to coherent nuclear wave-packet motion on excited-state potential-energy surfaces. This connection is well established experimentally: impulsive photo-excitation is known to launch vibrational wave packets whose phase and amplitude are directly tracked by femtosecond stimulated Raman spectroscopy (FSRS) and related impulsive Raman techniques \cite{dietze2016femtosecond}. The same nuclear coherences manifest as vibrational beat components that modulate electronic signals in 2DES \cite{milota2013vibronic,perlik2014distinguishing}. In this sense, the vibrational expectation values appearing in the present model provide a coarse-grained representation of experimentally accessible, phase-coherent nuclear motion dynamically coupled to electronic excitation. We also emphasise that the electronic--vibrational quantum correlations identified in this work might not be themselves direct experimental observables. Instead, they initiate and enable quantum synchronisation beyond the semi-classical factorised limit, which is experimentally accessible through oscillatory beatings, phase locking, and sustained population transfer under dissipation. The experimentally manifested quantity is therefore quantum synchronisation, which implicitly reflects underlying quantum correlations and, more importantly, accompanies throughout energy transfer dynamics.
\par 
Beyond the specific chromophore dimer studied here, the mechanism analysed has implications for studies of molecular and organic excitonic materials, in which the persistence of coherence and the resulting optical and transport properties are strongly shaped by vibronic coupling and interactions with the environment \cite{scholes2017using,bredas2004charge}. The environment-assisted transfer considered in this work arises from nonlinear interactions with environmental degrees of freedom, which indirectly mediate excitation transfer and stabilise electronic populations under continuous donor pumping and electronic dephasing \cite{chin2013role,anto2021capturing}. Importantly, this nonlinear mechanism emerges from quantum system--environment correlations and goes beyond linearised or semi-classical Markovian stochastic population dynamics \cite{sun2016non}. Related forms of such open quantum dynamics have been explored in collective phenomena in many-body material systems, such as magnon \cite{xu2025frohlich} and exciton--polariton condensation \cite{cherbunin2025quantum}, highlighting a broader context in which environment-mediated nonlinear interactions can lead to long-lived quantum coherence in nonequilibrium steady states. This setting motivates a concrete materials-level question: whether nonlinear modulation between system states by environmental coordinates can be sufficiently strong and controllable to sustain quantum coherence and efficient energy transport despite ambient dissipation \cite{cont25}. In the context of light-harvesting dimers, addressing this question would require moving beyond minimal exciton--vibrational dimer models to, for example, multi-chromophore assemblies, including energy funnels towards reaction centres with realistic parametrisation \cite{hestand2018expanded}. Such extensions also align with contemporary exciton modelling in molecular aggregates and organic semiconductors and provide a route for translating microscopic vibronic mechanisms into design principles for bio-inspired energy-transport architectures \cite{bredas2004charge}.

\nocite{*}

\begin{acknowledgments}
WX gratefully acknowledges Dr. Daniel R. Kattnig at the University of Exeter and Dr. Farhan T. Chowdhury at the Ohio State University for their helpful guidance and comments in manuscript preparation, as well as the financial support from the Erasmus+ traineeship grant.
\end{acknowledgments}

\bibliography{apssamp}

\end{document}